\documentclass[apjl]{emulateapj}
\usepackage{apjfonts}
\usepackage{mathptmx}

\slugcomment{accepted for publication in ApJL, 04/07/2006}

\shorttitle{NIR intraday variation of NGC4395}
\shortauthors{Minezaki et al.}

\begin{document}

\title{First Detection of Near-Infrared Intraday Variations in the Seyfert 1 Nucleus NGC4395}

\author{Takeo Minezaki\altaffilmark{1,2},
    Yuzuru Yoshii\altaffilmark{1,3}, Yukiyasu Kobayashi\altaffilmark{4},
    Keigo Enya\altaffilmark{5},
    Masahiro Suganuma\altaffilmark{4},
    Hiroyuki Tomita\altaffilmark{1},
    Shintaro Koshida\altaffilmark{6},
    Masahiro Yamauchi\altaffilmark{6},
    and Tsutomu Aoki\altaffilmark{7}
}

\altaffiltext{1}{Institute of Astronomy, School of Science, University of Tokyo,
    2-21-1 Osawa, Mitaka, Tokyo 181-0015, Japan}
\altaffiltext{2}{e-mail: minezaki@mtk.ioa.s.u-tokyo.ac.jp}
\altaffiltext{3}{Research Center for the Early Universe,
    School of Science, University of Tokyo,
    7-3-1 Hongo, Bunkyo-ku, Tokyo 113-0033, Japan}
\altaffiltext{4}{National Astronomical Observatory,
    2-21-1 Osawa, Mitaka, Tokyo 181-8588, Japan}
\altaffiltext{5}{Institute of Space and Astronautical Science,
    Japan Aerospace Exploration Agency,
    3-1-1, Yoshinodai, Sagamihara, Kanagawa, 229-8510, Japan}
\altaffiltext{6}{Department of Astronomy, School of Science, University of Tokyo,
    7-3-1 Hongo, Bunkyo-ku, Tokyo 113-0013, Japan}
\altaffiltext{7}{Kiso Observatory,
    Institute of Astronomy, School of Science, University of Tokyo,
    10762-30 Mitake, Kiso, Nagano 397-0101, Japan}

\begin{abstract}
We carried out a one-night optical $V$ and near-infrared $JHK$ 
monitoring observation of the least luminous Seyfert 1 galaxy, NGC4395, 
on 2004 May 1, and detected for the first time the intraday flux 
variations in the $J$ and $H$ bands, while such variation was not 
clearly seen for the $K$ band.  The detected $J$ and $H$ variations 
are synchronized with the flux variation in the $V$ band,
which indicates that the intraday-variable component of near-infrared 
continuum emission of the NGC4395 nucleus is an extension of power-law 
continuum emission to the near-infrared and originates in
an outer region of the central accretion disk.  On the other hand, from 
our regular program of long-term optical $BVI$ and near-infrared 
$JHK$ monitoring observation of NGC4395 from 2004 February 12 until 
2005 January 22, we found large flux variations in all the bands on 
time scales of days to months.  The optical $BVI$ variations are
almost synchronized with each other, but not completely with the near-infrared 
$JHK$ variations.  The color temperature of the near-infrared variable 
component is estimated to be $T=1320\sim 1710$ K, in agreement with 
thermal emission from hot dust tori in active galactic nuclei (AGNs).
We therefore conclude that the near-infrared variation consists of
two components having different time scales,
so that a small $K$-flux variation on a time 
scale of a few hours would possibly be veiled by large variation of 
thermal dust emission on a time scale of days. 
\end{abstract}



\section{Introduction}
The near-infrared continuum emission of type 1 AGNs
is considered to be dominated by thermal emission
from hot dust surrounding the central engine
\citep{kob93,glikman06}.
However, the contribution of the accretion disk
to the near-infrared continuum emission
is suggested by the standard accretion disk model
with large outer radius
that predicts the power-law continuum emission
of $f_{\nu }\propto \nu ^{+1/3}$
at long wavelengths \citep{ss73}.
If this is the case, part of the near-infrared emission
would originate in the outer accretion disk,
although it is difficult to separate the accretion disk component
from the dominant thermal dust component
by the simple method of spectral decomposition.
In order to clearly separate the accretion disk component,
techniques using variability \citep{tomita06} and
polarization \citep{kishimoto05}
were recently proposed and applied to some AGNs.

NGC4395 is a unique object known as
the least luminous Seyfert 1 galaxy \citep{filippenko89},
having a bolometric luminosity of
only $L_{\rm bol}\sim 10^{40-41}$ ergs s${}^{-1}$
with broad emission lines in the UV and optical spectra
\citep{filippenko89,filippenko93,kraemer99}.
Its spectral energy distribution (SED) from
the X-ray to radio wavelengths resembles those of
normal type 1 AGNs rather than those of
low-ionization nuclear emission-line regions (LINERs)
\citep{moran99}.
Although a compact radio source is detected,
it is a radio-quiet AGN \citep{houlvestad01,wrobel01,hopeng01}.
The mass of the central black hole measured by 
reverberation-mapping observation of the C {\small IV} emission line
is as small as $M_{\rm BH}=(3.6\pm1.1)\times 10^{5}\ M_{\sun }$
\citep{peterson05}.

Moreover, the NGC4395 nucleus shows
extreme variability at many wavelengths,
probably because of its low luminosity and small black hole mass.
Fast and large-amplitude variations were observed in the X-ray
\citep{lira99,iwasawa00,shih03,vaughan05,moran05},
and intraday variations were observed
in the UV and optical \citep{peterson05,skelton05}.
In the near-infrared,
while \citet{quillen00} reported
a 15\% flux variation in the $H$ band
at an interval of three weeks,
no detailed intraday monitoring observation 
has been carried out yet.

In this paper, we report the results
of the optical and near-infrared intraday monitoring observation
of the Seyfert 1 nucleus in NGC4395.
As observed in the X-ray, UV and optical,
we found the intraday variation in the $J$ and $H$ bands
almost synchronized with the optical intraday variation.
We conclude that the emission of
the intraday-variable component in the optical and near-infrared
originates in the central accretion disk.
Throughout this paper,
we adopt 4.0 Mpc for the distance of NGC4395 \citep{thim04}.


\section{Observations}
A one-night monitoring observation was carried out on 2004/05/01 (UTC),
while our regular program of long-term monitoring observation
started from 2004/02/12 (UTC),
using the multicolor imaging photometer (MIP) mounted on
the 2 m telescope
of the Multicolor Active Galactic Nuclei Monitoring
(MAGNUM) project 
at the Haleakala Observatories in Hawaii
\citep{kob98a,kob98b,yoshii02}.
Here, we describe the procedures of observation and data reduction
only briefly, because most of them have been
described in \citet{mine04}.

Our one-night monitoring observation
started at 06:31 and ended at 12:35 on 2004/05/01 (UTC).
Using the MIP's simultaneous
optical and near-infrared imaging capability,
a sequence of $(V,K)$, $(V,H)$, and $(V,J)$ imagings
was repeated cyclically during the observation,
and 59 data points for $V$ and
20 data points for each of $J$, $H$, and $K$ were obtained.
The average monitoring interval was
six minutes for $V$ and 18 minutes for $J$, $H$, and $K$.
In addition, the long-term monitoring observation
started from 2004/02/12 (UTC)
and 11 data points were obtained until 2005/01/22 (UTC).
After 2004/04/29 (UTC),
the observation each night
was carried out quasi-simultaneously
(within 10 minutes)
in the $B$, $V$, $I$, $J$, $H$, and $K$ bands.

The nuclear flux was measured relative to a nearby reference star
($\alpha=12^{\mathrm h} 25^{\mathrm m} 50.91^{\mathrm s}$,
$\delta=+33\arcdeg 33\arcmin 10.1\arcsec$ (J2000)),
which was observed along with the target,
within the field of view of the MIP.
The optical fluxes of the star were calibrated 
with respect to a photometric standard star in \citet{landolt92},
and the near-infrared fluxes were taken
from the 2MASS All-Sky Catalog of Point Sources \citep{cutri03}.
The diameter of photometric aperture was
2.8 arcsec for the optical, and 2.4 arcsec for the near-infrared.
The flux of the host galaxy within the aperture
was not subtracted.


\section{Results}
\begin{figure}
\plotone{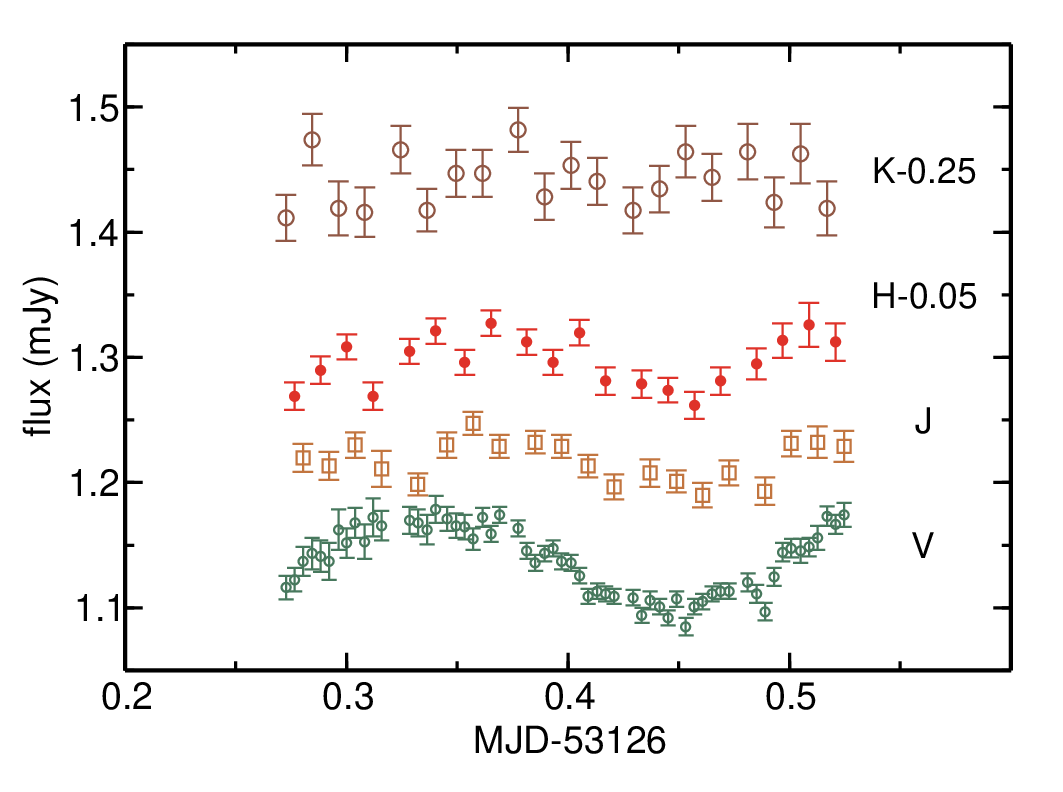}
\caption{
The optical and near-infrared light curves
of the NGC4395 nucleus on 2004/05/01 (UTC);
$V$ (small open circles), $J$ (open squares),
$H$ (small filled circles),
and $K$ (open circles).
For clarity, an offset of $-0.05$ mJy is applied to $H$,
and $-0.25$ mJy to $K$.
\label{fig1}}
\end{figure}

\begin{figure}
\plotone{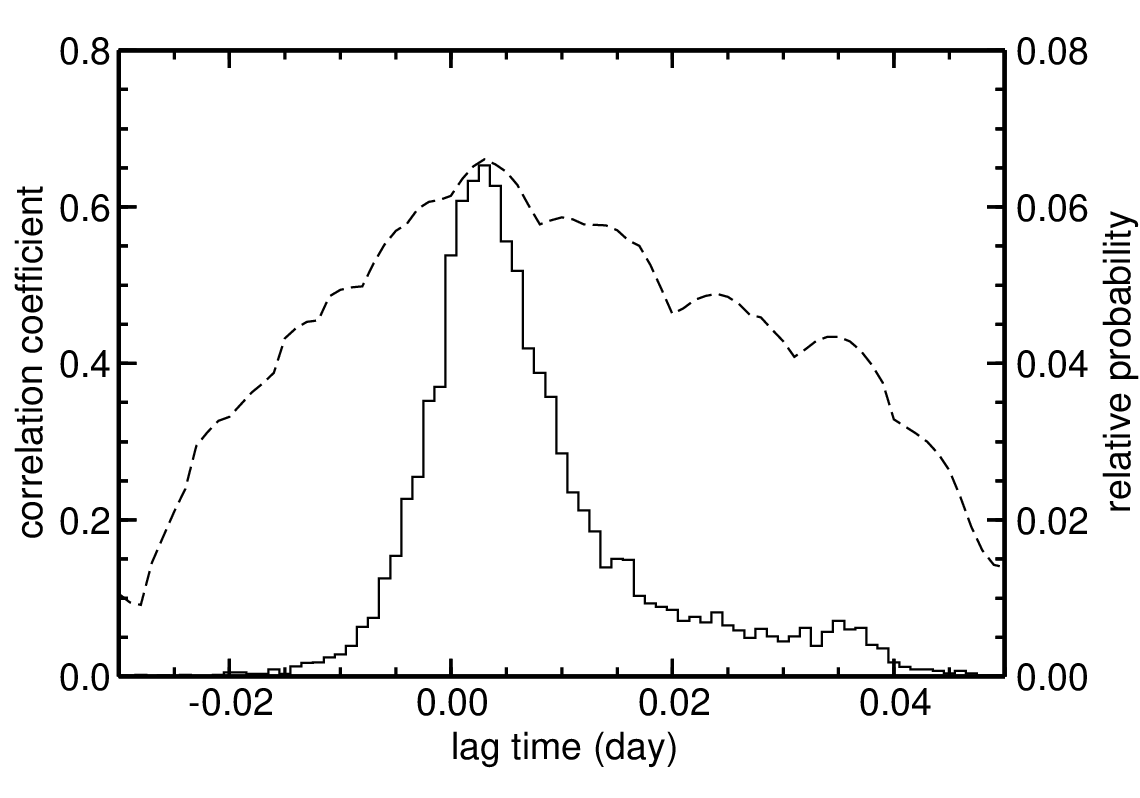}
\caption{
The CCF (dashed line) and the CCCD (solid line)
between the $V$-band and the $J+H$ combined light curves
on 2004/05/01 (UTC).
The CCF is peaked at $0.003$ days and 
the CCF centroid is estimated
as $\tau _{\rm cent}= 0.005^{-0.006}_{+0.011}$ days
or $7.2^{-8.6}_{+15.8}$ minutes
from the CCCD.
\label{fig2}}
\end{figure}

Figure \ref{fig1} shows
the optical and near-infrared light curves
of the NGC4395 nucleus on 2004/05/01 (UTC).
Galactic extinction was corrected for
according to \citet{schlegel98}.
Clear flux variations in the $V$, $J$, and $H$ bands
are found during the monitoring period of six hours,
and this is the first detection of intraday variation
in the near-infrared for normal Seyfert 1 galaxies.
The amplitude of variations was $0.05-0.09$ mag,
however,
the real amplitude for the NGC4395 nucleus must be larger,
because non-variable fluxes
such as the host galaxy flux and the narrow-line flux
were not subtracted from the data.
On the other hand, intraday variation
is not clearly seen in the $K$ band.

Apparently, the intraday variations in the $V$, $J$, and $H$ bands
are almost synchronized.
In order to estimate a possible lag
between the optical and near-infrared intraday variations,
we applied a cross-correlation analysis.
First, 
in order to increase the number of data points,
the $J$-band and the $H$-band light curves were combined
according to the linear regressions
of the $J$- and $H$-band fluxes to the $V$-band flux
obtained simultaneously during
the one-night monitoring observation.
Then a cross-correlation function (CCF)
was computed based on the linear interpolation method
\citep{gp87,wp94}.
Since the number of $V$-band data points is larger
and their photometric accuracy is better,
the $V$-band light curve was interpolated.
The lag of the CCF centroid $\tau _{\rm cent}$ was adopted
to represent the lag between two light curves.
The error of $\tau _{\rm cent}$
was estimated by a model-independent Monte Carlo simulation
called the ``flux randomization/random subset selection'' (FR/RSS) method
\citep{peterson04}.
The cross-correlation centroid distribution (CCCD)
was made based on the simulation of 10000 realizations.
Figure \ref{fig2} shows the CCF and the CCCD
between the $V$-band and the combined near-infrared light curves
on 2004/05/01 (UTC).
The near-infrared lag time behind $V$ 
is estimated from the CCCD
as $\tau _{\rm cent}= 0.005^{-0.006}_{+0.011}$ days,
or $7.2^{-8.6}_{+15.8}$ minutes.
The intraday variations in the near-infrared ($J$, $H$)
are almost synchronized with that in the optical
within an accuracy of $\sim 10$ minutes.
The short time scale of variation (a few hours)
and the synchronization to the optical variation ($\sim 10$ minutes)
suggest that
the intraday-variable component of flux in the near-infrared
originates in the central accretion disk.
That is, it is an extension of
optical power-law continuum emission to near-infrared wavelength.

\begin{figure}
\plotone{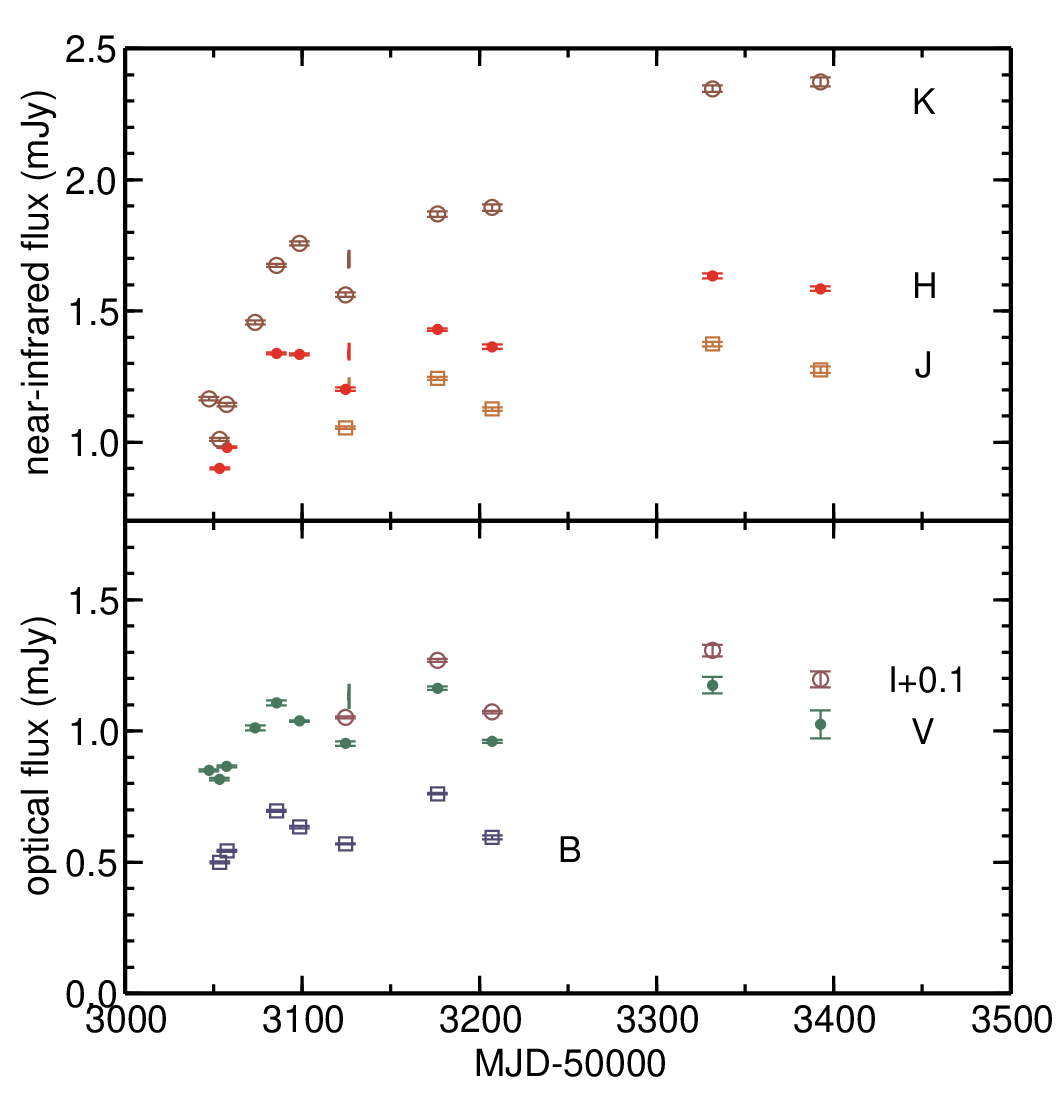}
\caption{
The optical and near-infrared light curves of the NGC4395 nucleus;
$B$ (open squares), $V$ (small filled circles),
and $I$ (open circles) in the lower panel,
and $J$ (open squares), $H$ (small filled circles),
and $K$ (open circles) in the upper panel.
An offset of $0.1$ mJy is applied to $I$ for clarity. 
The intraday variation data of 2004/05/01 (UTC)
are presented by vertical bars at MJD$=53126$.
The optical data at the last two epochs were noisy
because of instrumental problems,
and their $B$-band data were omitted.
\label{fig3}}
\end{figure}

\begin{figure}
\plotone{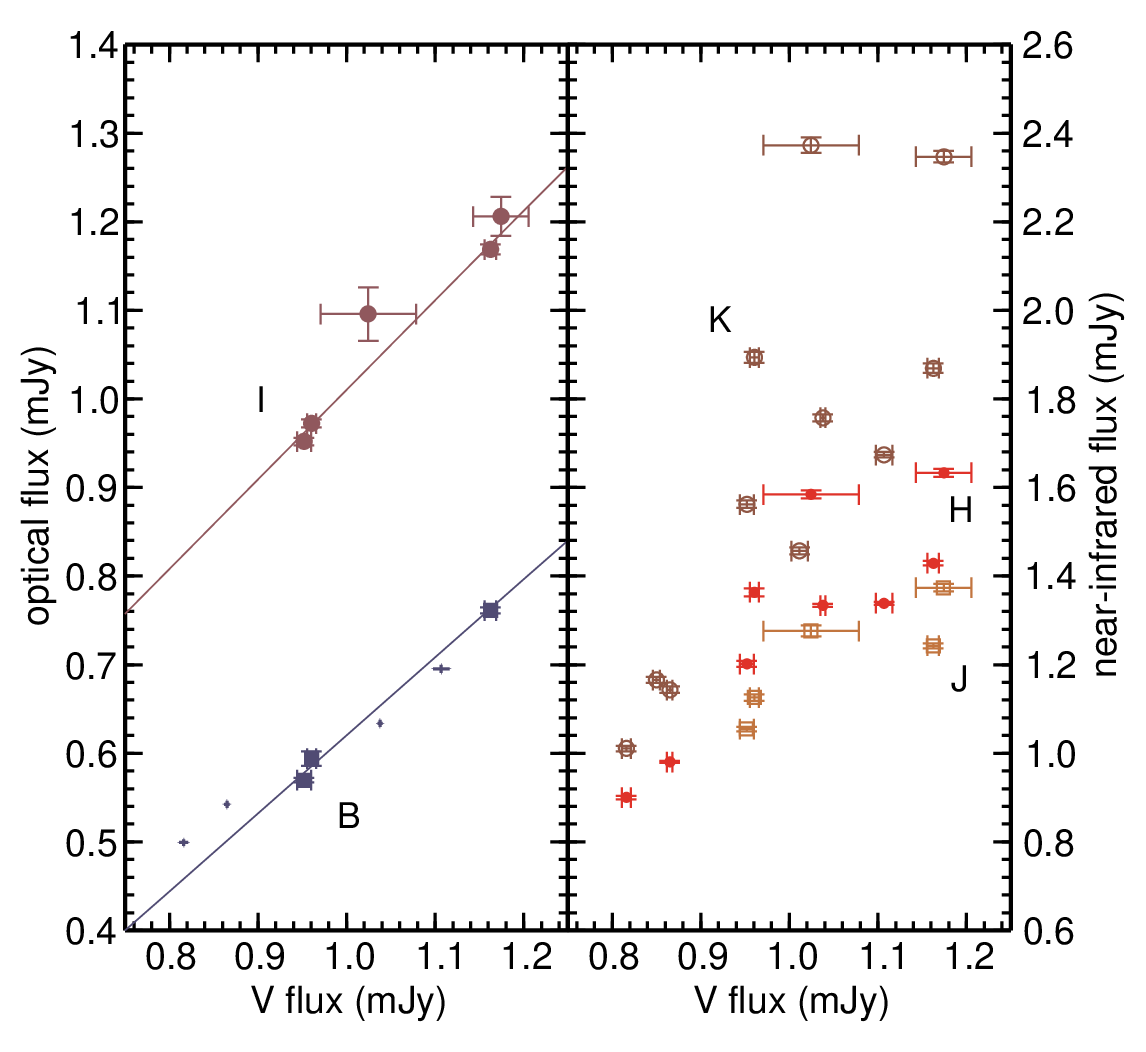}
\caption{
The flux-flux diagrams from the long-term monitoring of the NGC4395 nucleus.
Plotted against $V$ are the data for
$B$ (filled squares and crosses)
and $I$ (filled circles) in the left panel,
and $J$ (open squares), $H$ (small filled circles),
and $K$ (open circles) in the right panel.
The straight lines in the left panel are the regression lines
based on the $BVI$ data (filled symbols)
which were taken almost simultaneously within 10 minutes.
\label{fig4}}
\end{figure}

Figure \ref{fig3} shows the optical and near-infrared
light curves of the long-term monitoring observation
of the NGC4395 nucleus.
The optical and near-infrared fluxes are
clearly variable on time scales of days to months,
and the apparent amplitude of variation during
the monitoring period reaches
about a few tenths to one magnitude.
The optical flux variations in the $B$, $V$, and $I$ bands
are synchronized with each other,
while the near-infrared flux variations, especially in the $K$ band,
are not completely synchronized with those in the optical.

Figure \ref{fig4} shows the flux-flux diagrams
from the long-term monitoring observation.
The variations in the $B$ or $I$ bands
relative to the $V$ band are almost on straight lines,
which suggests that
the spectral shape of variable components
in the optical region stays constant
as reported for Seyfert galaxies \citep{winkler92,winkler97}.
The slope of these lines,
or the flux variation gradient (FVG),
corresponds to the power-law index of variable component.
From the regression analysis using
the $B$, $V$, and $I$ data points of the quasi-simultaneous observation,
we obtain $\alpha_{\nu }=-0.6\pm 0.2$ for $B$ to $V$,
and $\alpha_{\nu }=0.0\pm 0.1$ for $V$ to $I$,
which are similar to those of more luminous Seyfert galaxies
\citep{winkler97}.

In contrast,
the near-infrared $J$, $H$, and $K$ variations
largely deviate from straight lines,
and the deviation is larger for longer wavelengths,
which suggests that another variable component exists in the near-infrared.
We extracted the SED of this component
by evaluating the near-infrared flux difference
at two epochs for which the optical fluxes were almost the same.
The flux difference between 2004/04/29 and 2004/07/21 is
$\Delta f$ (mJy) $=0.025\pm0.009$ $(B)$,
$0.009\pm0.010$ $(V)$,
$0.020\pm0.006$ $(I)$,
$0.071\pm0.009$ $(J)$,
$0.162\pm0.011$ $(H)$,
and $0.332\pm0.015$ $(K)$.
The color temperature assuming the black body radiation
is estimated as $1710\pm 150$ K from $\Delta f_H/\Delta f_J$,
and $1320\pm 80$ K from $\Delta f_K/\Delta f_H$.
These estimates are consistent with
the near-infrared SEDs of QSOs
originating in the hot dust tori \citep{kob93,glikman06}.


\section{Discussion}
The near-infrared intraday variation of the NGC4395 nucleus
and its synchronization to the optical suggest
that there exists a near-infrared emitting region in the accretion disk,
so that the lag between the optical
and the near-infrared intraday variations
would be a direct measure of the extent of the accretion disk,
if the reprocessing of incident flux variation
to the near-infrared can be applied.
Accordingly,
a wavelength-dependent lag
in variations of power-law continuum emission is expected,
interpreted in terms of light-travel time
from the central energy source to
each radius of the accretion disk
\citep{collier99}.

Therefore, 
the outer radius of the accretion disk
indicated by the possible lag between the $V$-band and
the near-infrared intraday variations
is worthwhile comparing with
the size of the broad emission-line region (BLR).
Our measurement of near-infrared lag time behind $V$
is $\tau_{\rm cent} = 0.005^{-0.006}_{+0.011}$ days (2004/05/01).
\citet{peterson05}
reported from the reverberation mapping program
for the C {\small IV} emission line of the NGC4395 nucleus
that the C {\small IV} lag time behind UV continuum $(\lambda=1350$ \AA$)$
is $\tau_{\rm CIV} =0.033^{+0.017}_{-0.013}$ days (2004/04/10)
and $\tau_{\rm CIV} =0.046^{+0.017}_{-0.013}$ days (2004/07/03).
Apparently,
$\tau_{\rm cent}$ is smaller than $\tau_{\rm CIV}$.
If we adopt the wavelength-dependent relation
$\tau (\lambda )\propto \lambda ^{4/3}$
given by \citet{collier99},
the near-infrared lag behind the UV continuum
would be $\sim 1.3-1.4$ times larger
than that behind $V$.
In addition,
the inner region of the BLR of the NGC4395 nucleus
would be a few times smaller than
the C {\small IV} emitting region
because the BLR of Seyfert galaxies is stratified \citep{peterson99}.
These considerations suggest that
the outer radius of the accretion disk
might be comparable to the inner radius of the BLR
in the NGC4395 nucleus.

Next, we discuss the contribution of
the power-law continuum component in the $K$ band.
As described in the previous section,
thermal emission from the dust torus
largely contributes to the $K$-band flux
of the NGC4395 nucleus.
Since the dust torus is expected to be extended,
the short time scale variation of incident flux
is smeared out and the thermal flux would not be
variable on such a time scale.
Then, even if there was an intraday flux variation of
the power-law continuum component in the $K$ band,
it would be veiled by
the large amount of thermal flux from the dust torus.
We estimated an average of $K$-band fluxes
at peaks (five data points) and at valleys (six data points)
of the $V$-band intraday variation,
and derived their difference.
The $3\sigma$ upper limit of the flux difference of $0.04$ mJy
represents an amount of intraday variation in the $K$ band.
On the other hand,
such a difference in the $J$ and $H$ bands is about 0.03 mJy.
Therefore,
it is possible to consider that
an intraday flux variation in the $K$ band is comparable to
those in the $J$ and $H$ bands.

Finally,
we estimate the inner radius of the dust torus
according to the correlation
between the $V$-band absolute magnitude
and the lag time of $K$-band flux variation
reported by \citet{mine04}.
The host galaxy flux within the aperture
is estimated as $0.26$ mJy
from the $I$-band magnitude of
the central star cluster in NGC4395 \citep{filippenko03}
and by assuming typical $V-I$ color of Scd galaxies
\citep{fukugita95}.
The flux of the [O {\small III}] narrow line
was taken from \citet{skelton05},
and its contamination of the $V$-band flux
is estimated as $0.24$ mJy.
After subtracting these fluxes,
the $V$-band absolute magnitude of the NGC4395 nucleus
is estimated as $M_V\sim -11$ mag,
which leads to a lag time of $\tau _K\sim 1$ day.
The thermal flux from such a dust torus,
whose inner radius is $\sim 1$ light-day,
would not be variable on a time scale of hours,
veiling the possible intraday variation
of the power-law continuum component in the $K$ band.


The present paper indicates that more intense multi-wavelength monitoring 
observation of the NGC4395 nucleus every 10 minutes a night will definitely 
extract the synchronized intraday variation of power-law continuum component 
up to the near-infrared $K$ band, and a wavelength-dependent lag behind 
X-ray or UV is able to probe the radial structure of central accretion disk.  
This intense multi-wavelength monitoring observation over a month will 
alternatively extract the near-infrared $K$ variation of thermal dust 
component, and its lag behind UV or optical gives an accurate measure of 
the inner radius of minimum dust torus in the least luminous nucleus among 
Seyfert 1 AGNs.  These lag measurements of greater accuracy, together with the 
C {\small IV} lag \citep{peterson05}, will finally uncover a full range of 
radial structure consisting of the accretion disk, the BLR, and the dust 
torus in the NGC4395 nucleus.

\acknowledgments

 We thank staff at the Haleakala Observatories
 for their help with facility maintenance.
 This research has been supported partly by
 the Grant-in-Aid of Scientific Research
 (10041110, 10304014, 11740120, 12640233, 14047206, 14253001, 14540223, 16740106)
 and COE Research (07CE2002) of
 the Ministry of Education, Science, Culture and Sports of Japan.

\end{document}